\documentclass[letterpaper,twocolumn,10pt]{article}
\usepackage{graphics}
\usepackage{endnotes}
\usepackage{usenix}

\usepackage{url}
\usepackage{verbatim}
\usepackage{amssymb}

\newcommand{\eat}[1]{}

\begin{document}

\date{}
\title{\Large \bf Requirements for Digital Preservation Systems: A Bottom-Up Approach}

\author{
{\rm David S. H. Rosenthal}
\and
{\rm Thomas Robertson}
\and
{\rm Tom Lipkis}
\and
{\rm Vicky Reich}
\and
{\rm Seth Morabito}\\
Stanford University Libraries, {CA}
}

\maketitle


\subsection*{Abstract}

The field of digital preservation is being defined by a set of
standards developed top-down,  starting with an abstract reference
model (OAIS) and gradually adding more specific detail.  Systems
claiming conformance to these standards are entering production
use.  Work is underway to certify that systems conform to
requirements derived from OAIS.

We complement these requirements derived top-down by presenting an
alternate,
bottom-up view of the field.
The fundamental goal of these systems is to ensure that the information
they contain remains accessible for the long term.  
We develop a parallel set of requirements based on observations of how
existing systems handle this task,
and on an analysis of the threats to achieving the goal.
On this basis we suggest \emph{disclosures} that systems should provide
as to how they satisfy their goals.

{\tiny\verb!$Revision: 1.31 $!}

\section{Introduction}
\label{sec:introduction}

The field of digital preservation systems has been defined by the
Open Archival Information System (OAIS) standard
ISO\ 14721:2003~\cite{OAISstandard},  which provides
a high-level reference model.  This model has been very useful.
It identifies the participants,  describes their roles and
responsibilities,  and classifies the types of information
they exchange.  However,  because it is only a high-level reference
model,  almost any system capable of storing and retrieving
data can make a plausible case that it satisfies the OAIS
conformance requirements.

Work is under way to elaborate the OAIS reference model
with sufficient detail to allow systems to be certified by an ISO\ 9000-like
process~\cite{RLGCert2005},
and to allow systems to inter-operate on the basis of common specifications
for ingesting and disseminating information~\cite{vandeSompel2004, bekaert2005}.
In the same way that
ISO\ 14721 was developed top-down,  these efforts are also top-down.

Several digital preservation systems are in, or about to
enter, production use preserving content society deems important.
It seems an opportune moment to complement the OAIS top-down
effort to generate requirements for such systems with a bottom-up
approach.
We start by identifying the goal the systems are intended to achieve,
and then analyze the spectrum of threats that might prevent them doing so.
We list the strategies that systems can adopt to counter these threats,
providing examples from some current systems showing how the
strategies can be implemented.
We observe that current systems vary in the set of threats they
consider important,  in the strategies they choose to implement,  and
in the ways in which they implement them.  Given this,  and the
relatively short experience base on which to draw conclusions as to
which approaches work better,  we agree with the top-down proponents
that setting requirements explicitly or implicitly
mandating specific technical approaches seems imprudent.
This paper presents a list of \emph{disclosures} that we suggest should form
part of the basis for comparing and certifying systems in the medium term.

We draw on our six years experience developing, deploying,  and talking
to librarians and publishers about the
LOCKSS~\footnote{LOCKSS is a trademark of
Stanford University,  It stands for Lots Of Copies Keep Stuff Safe}
digital preservation system~\cite{Maniatis2005ToCS}.  Our descriptions
of other systems are based on published materials and past discussions
with their implementors.
Although a comprehensive survey of current digital preservation
systems would be useful,  this paper does not attempt such a survey.
We refer to systems simply to demonstrate that particular techniques
are currently in use and do not attempt to list all systems using
them.  Note that we believe all the systems to which we refer satisfy
the conformance requirements of ISO\ 14721.

\section{Goal}
\label{sec:Goal}

The goal of a digital preservation system is that \emph{the information
it contains remains accessible to users over a long period of time}.

The key problem in the design of such systems is that the period
of time is very long,
much longer than the lifetime of individual storage media,
hardware and software components, and the formats in which the
information is encoded.
If the period were shorter,
it would be simple to satisfy the requirement by storing the information
on suitably long-lived media embedded in a system of similarly
long-lived hardware and software.

No media, hardware or software exists in whose longevity
designers can place such confidence.
They must therefore anticipate failures
and obsolescence,  designing systems with three key properties:
\begin{itemize}
\item At minimum, the system must have no single point of failure;
it must tolerate the failure of any individual component
(see Section~\ref{sec:Strategies:replication}).
In general, systems should be designed to tolerate more than one
simultaneous failure.
\item Media,  software and hardware must flow through the system
over time as they fail or become obsolete,  and are replaced.
The system must support diversity among its components
to avoid monoculture vulnerabilities,
to allow for incremental replacement,
and to avoid vendor lock-in.
(see Section~\ref{sec:Strategies:diversity}).
\item Most data items in an archive are accessed infrequently.  A system that
detected errors and failures only upon user access would
be vulnerable to an accumulation of latent errors\cite{KariThesis}.
The system must provide
for regular audits at intervals frequent enough to keep the
probability of failure at acceptable levels (See
Section~\ref{sec:Strategies:audit}).
\end{itemize}

The major contrast between the top-down and bottom-up approaches can
be summed up as being between the figure and the ground in a view of the system.
The top-down approach naturally focuses on what the system \emph{should}
do,  in terms of exchanging this kind of data and this kind of meta-data
with these types of participant.  Whereas the bottom-up approach
naturally focuses on what the system \emph{should not} do,  in terms
of losing data or delaying access under specific types of failures.
Both views have value to system designers.

\section{Threats}
\label{sec:Threats}

We concur with the recent National Research Council recommendations
to the National Archives~\cite{Sproull2005} that
the designers of a digital preservation system need a clear vision
of the threats against which they are being asked to protect their
system's contents,  and those threats under which it is acceptable
for preservation to fail.  Note that many of these threats are not
unique to digital preservation systems,  but their specific mission
and very long time horizons incline such systems to view the threats
differently from more conventional systems.

To assist in the development of these
threat models,  we present the following taxonomy of threats.
Threat models should either include or
explicitly exclude at least these threats:

\begin{description}
\item [Media Failure] All storage media must be expected to degrade with
time,  causing irrecoverable bit errors,  and to be subject to sudden
catastrophic irrecoverable loss of bulk data such as
disk crashes~\cite{Talagala1999} or loss of off-line media~\cite{Reuters2005}.
\item [Hardware Failure] All hardware components must be expected to
suffer transient recoverable failures,  such as power loss,  and
catastrophic irrecoverable failures, such as burnt-out power supplies.
\item [Software Failure] All software components must be expected to suffer
from bugs that pose a risk to the stored data.
\item [Communication Errors] Systems cannot assume that the network transfers
they use to ingest or disseminate content
will either succeed or fail within a specified time period,
or will actually deliver the content unaltered.
A recent study ``suggests that between one (data) packet in every 16 million
packets and one packet in 10 billion packets will have an undetected checksum
error''~\cite{stone00when}.
\item [Failure of Network Services] Systems must anticipate that the external
network services they use,  including resolvers such as those for
domain names~\cite{RFC1034} and persistent URLs~\cite{OCLCPURL},
will suffer both transient and irrecoverable failures both of the
network services and of individual entries in them.
As examples,  domain names will
vanish or be reassigned if the registrant fails to pay the registrar,
and a persistent URL will fail to resolve if the resolver service
fails to preserve its data with as much care as the digital preservation
service.
\item [Media \& Hardware Obsolescence] All media and hardware components
will eventually fail.  Before that, they may become obsolete in the sense of no
longer being capable of communicating with other system components or
being replaced when they do fail.  This problem is particularly acute
for removable media,  which have a long history of remaining theoretically
readable if only a suitable reader could be found~\cite{Keeton2001}.
\item [Software Obsolescence] Similarly,  software components will become
obsolete.  This will often be manifested as \emph{format obsolescence} when,
although the bits in which some data was encoded remain accessible,
the information can no longer be decoded from the storage format into a
legible form.
\item [Operator Error] Operator actions must be
expected to include both recoverable and irrecoverable errors.
This applies not merely to the digital preservation application itself,
but also to the operating system on which it is running,  the other
applications sharing the same environment,  the hardware underlying them,
and the network through which they communicate.
\item [Natural Disaster] Natural disasters, such as
flood~\cite{StanfordLibraryFlood}, fire and earthquake must be anticipated.
They will typically be manifested by other types of threat,  such as
media,  hardware and infrastructure failures.
\item [External Attack] Paper libraries and archives are subject to
malicious attack~\cite{BookMutilation};  there is no reason to expect their
digital equivalents to be exempt.  Worse,  all systems connected to
public networks are vulnerable to viruses and worms.
Digital preservation systems must either
defend against the inevitable attacks,  or be
be completely isolated from external networks.
\item [Internal Attack] Much abuse of computer systems involves insiders,
those who have or used to have authorized access to the
system~\cite{Keeney2005}.
Even if a digital
preservation system is completely isolated from external networks,
it must anticipate insider abuse.
\item [Economic Failure] Information in digital form is much more vulnerable
to interruptions in the money supply than information on paper.  There are 
ongoing costs
for power,  cooling,  bandwidth, system administration, domain registration,
and so on.  Budgets
for digital preservation must be expected to vary up and down,
possibly even to zero,  over time.
\item [Organizational Failure] The system view of digital preservation must
include not merely the technology but the organization in which it is
embedded.  These organizations may die out,  perhaps through bankruptcy,
or change missions.  This may deprive the digital preservation technology
of the support it needs to survive.  System planning must envisage the
possibility of the asset represented by the preserved content being
transferred to a successor organization,  or otherwise being properly
disposed of.
\end{description}

For each of these types of failure,  it is necessary to trade off 
the cost of defense against the level of system degradation under the
threat that is regarded as acceptable for that cost.  The degradation
may be evaluated in terms of the following questions:
\begin{itemize}
\item What fraction of the system's content is irrecoverably lost?
\item What fraction of the user population suffers what delay in
accessing the impaired but recoverable fraction of the system's
content?
\end{itemize}

Designers should be aware that these threats are likely to
be \emph{highly correlated}.  For example,  operators stressed by
responding to one threat,  such as hardware failure or natural
disaster,  are far more likely to make mistakes than they are
when things are calm~\cite{Reason1990short}.  Equally,  software
failures are likely to be triggered by hardware failures which
present the software with conditions its designers failed to
anticipate and under which it has never been tested.
Mean Time Between Failure estimates are typically based
on the assumption that failures occur independently
(e.g.~\cite{Patterson1988});  even small
correlations between the failures can render the estimates wildly
optimistic.

\section{Strategies}
\label{sec:Strategies}

We now survey the strategies that system designers can employ to
survive these threats.

\subsection{Replication}
\label{sec:Strategies:replication}

The most basic strategy exploits the fundamental attribute that
distinguishes digital from analog information,  the possibility of copying it
without loss of information,  to store multiple replicas of the information
to be preserved.  Clearly,  a single replica subject to the threats above
has a low probability of long-term survival,  so replication is
a \emph{necessary} attribute of a digital preservation system but it
is far from \emph{sufficient},  as anyone who has had trouble restoring
a file from a backup copy can appreciate.

Examples of a common approach to replication among current digital
preservation systems are Florida's DAITSS~\cite{DAITSS2004} and the
system under development at the British Library (BL)~\cite{Baker2005a}.
Both use a fixed number (3,4) of replicas,  automatically creating
replicas of each submitted item at geographically distributed
sites.  Each site may create further,  off-line backup replicas.

An example of a system using dynamic, and much higher levels,  of replication
is the LOCKSS peer-to-peer digital preservation system, in which each
participating library collects its own copy of the
information in which it is interested.  The level of replication for an item is
set by the number of libraries that collect it,  which ranges in
the deployed system from 6 to 80 or more.  The LOCKSS auditing process
(see Section~\ref{sec:Strategies:audit}) advises peer operators to establish
more replicas when the number
their peer can locate drops below a preset threshold.

\subsection{Migration}
\label{sec:Strategies:migration}

The creation and management of replicas which lies at the base of a digital
preservation system involves a process of migration,  between instances of
the same type of storage medium,  from one medium to another,  and from
one format to another.  Migrations can be exceptional events,  handled
by the system operators perhaps on a batch basis,  or routine events,  handled
automatically by the system without operator intervention.

Migration between instances of the same medium,  for example network
transfers from mass storage at one site to mass storage at another,
is typically used to implement replication and to refresh media.
All systems employing
replication appear to use it.  It can be effective against media and
hardware failures.

The classic example of migration between media is tape backup,  used by
many systems.  It can be effective against media,  hardware and software
failures and obsolescence.

Format migration can be an effective strategy to combat software obsolescence.
Many systems do format migration,  in some cases preemptively on
ingress (e.g. ANA, the Australian National Archives~\cite{Heslop2002}),
in some cases on a batch basis (e.g. DAITSS) and in some cases
on access (e.g. LOCKSS~\cite{Rosenthal2005a}).
Preemptive migration on input is often called \emph{normalization}.
All such systems of which we are aware plan to preserve the original
format;
some in addition preserve the result of format migration.

Some systems,  for example that at the Koninklijke Bibliotheek (KB,
National Library of the Netherlands),
avoid the issue of format migration by accepting information for preservation
only in formats believed to be suitable,
in KB's case PDF~\cite{Wijngaarden2004},
and by pursuing a strategy of emulation~\cite{Lorie2004,Rothenberg1995} to
ensure that the interpreters for the chosen formats will remain usable.

\subsection{Transparency}
\label{sec:Strategies:transparency}

Digital preservation technology shares some attributes with encryption
technology.  Perhaps the most important is that in both cases the
customer has no way to be sure that the system will continue to perform its
assigned task,  of preserving or preventing access to the system's
content as the case may be.  An encryption system may be broken or
misused and therefore reveal content.  However long you watch a
digital preservation system,  you can never be sure it will continue
to provide access in the future.

In both cases transparency is key to the customer's confidence in the
system.  Just as open source,  open protocols and open interfaces
provide the basis for the public review that allows customers
to have confidence in encryption systems such as AES~\cite{AEStandardProcess},
similar reviews based on similar introspection are needed if customers
are to have confidence that their digital preservation systems
will succeed.  Examples of open-source digital preservation systems
include the LOCKSS system and MIT's DSpace
system~\cite{MacKenzieSmith2004}.

An essential precaution against the software of a digital preservation
system becoming obsolete is that it be preserved with at least as
much care as the information which it is preserving.  Open source
makes this easy.
Open protocols and open interfaces are a necessary but not sufficient
precondition for diverse implementations of system components (see
Section~\ref{sec:Strategies:diversity}) and for effective
``third-party'' audit mechanisms
(see Section~\ref{sec:Strategies:audit:external}).
We have yet to see examples of diverse implementations or third-party
audit in practice.

Transparency is also the key to the ability to perform format
migration.  Widely used data formats well-supported by open
source interpreters,  such as the majority of those used on the
Web,  are easy to migrate~\cite{Rosenthal2005a,Quiggle2005}.  Proprietary
formats,  particularly those supported by a business model that
thrives on backwards incompatibility,  are much harder.
The hardest of all are proprietary formats entwined with proprietary
hardware,  such as game consoles.

While there is clearly a role in digital preservation for proprietary,
closed software that implements open interfaces or formats,  using closed
software and proprietary interfaces or formats renders the preserved information
hostage to the vendor's survival and is hard to justify.
Transparency in general,  and open source in particular, can be an
effective strategy against all forms of obsolescence.
Access to the source encourages wide review of the system for vulnerabilities,
which can help prevent attacks succeeding.
Open source can also be effective against economic
failure,  by preventing an organization's financial troubles
from dooming the system's technology.  Open source software
will not be viewed as an asset to be sold or 
fall under the control of a bankruptcy court.
Note that open source software may be supported by major companies such
as IBM and Sun Microsystems, rather than depending upon volunteer
community support.

Despite the best efforts of system designers and implementors,
and despite the certifications expected to be available for
digital preservation systems,  data will be lost.
To improve the performance of systems over time,  it is essential
that lessons be learned from incidents that risk or cause data loss.
We can expect that such incidents will be infrequent,  making
it important to extract the maximum benefit from each.
Past incidents suggest that an institution's reaction to data loss
is typically to cover it up,  preventing the lessons being learned.
This paper shows this problem,
in that we have no way to cite or discuss the details of several
incidents of this kind known to practitioners via the grapevine.

This problem is familiar in aviation safety,  and it has led
NASA to establish the Aviation Safety Reporting System~\cite{ASRS}.
Through this no-fault reporting system incidents can be reported
in suitably anonymized form,  allowing the community to learn from
rare but important failures without penalizing those reporting them.
A similar system would be of great benefit in digital preservation.

\subsection{Diversity}
\label{sec:Strategies:diversity}

Systems lacking diversity,  in the extreme \emph{mono-cultures},  are
vulnerable to catastrophic failure.  Ideally,  a digital preservation
system should provide diversity at all levels,  but most systems
provide it at only a few, citing cost considerations:
\begin{itemize}
\item Most systems use off-line media to provide diversity in
\emph{media} for storing replicas,  and to isolate some
replicas as far as possible from network-borne threats.
\item Many systems use \emph{geographic} dispersion of on-line replicas to
counter threats of natural disaster (e.g. DAITSS and the BL's system).
Most systems using off-line backups store them off-site,  again
providing geographic diversity.  The LOCKSS system has replicas
scattered around the world.
\item The BL's system is an example of explicit planning for diversity
in \emph{hardware and vendors} to support a process of ``rolling
procurement'' and
``rolling replacement''~\cite{Baker2005a}.  The library's continuous
collection program means that the system must grow incrementally,
its availability requirements mean that replicas must be replaced
incrementally (a sound approach to preventing correlated administration
errors~\cite{Lettice2004}),
and its long planned lifetime means that vendor lock-in is
unacceptable.
\item Similar considerations apply to \emph{software}.  There should
be a diversity of software among the replicas.  The BL's system
anticipates that at any one time different replicas will be running
earlier or later versions of their management software,  and
that the different manufacturers of the underlying storage
technologies will provide some level of software diversity.
\item The BL's and LOCKSS systems are examples of diversity of
\emph{system administration}.
Each replica is independently administered;
there is no single password whose compromise could affect all replicas.
Given the prevalence of human error and insider abuse of computer systems,
unified system administration should be an unacceptable feature
of digital preservation.
\item The Portico~\cite{Waters2005} and LOCKSS systems are striving for
diversity of \emph{funding}.  As regards the
peers actually storing content the LOCKSS system is already diverse;
each peer is owned
and supported by its host library so no single budget cut or
administrative decision can cause the system as a whole to lose content.
Portico as a whole and the team that supports the LOCKSS system
are both in the process of transition
from sole-source grant funding,  to support
by the libraries using the service.  In this model no single
budget decision would affect more than a few percent of the team's
total income.
\end{itemize}

The risk of inadequate diversity is particularly acute
for networked computer systems such as digital preservation
systems.  Techniques have been available for some years
by which an attacker can compromise almost all systems sharing
a vulnerability in a very short time~\cite{Staniford2002}.
Worms such as Slammer~\cite{SlammerAnalysis} have used them in the wild.
System designers would be unwise to believe that they can
construct, configure, upgrade, and expand systems for the long term
that are not exploitable in this way.

Replicated systems can prevent attacks resulting in catastrophic
failure by arranging that replicas do
not share common implementations and thus common vulnerabilities.
This approach has
been explored in a data storage context at UCSD~\cite{Junqueira2005},
but we are not aware of any production digital preservation systems currently
using diversity in a similar way.  The LOCKSS system is taking its first
steps in this direction;  a version of its network
appliance~\cite{Rosenthal2003b} based on a second operating system
is under development.

\subsection{Audit}
\label{sec:Strategies:audit}

Most data items in digital preservation systems are,  by their archival
nature,  rarely accessed by users.  Although in aggregate systems
such as the Internet Archive may satisfy a large demand from users,
the average interval between successive accesses to any individual item in
the archive is long~\cite{Saltzer1990}.
Many systems (e.g. DAITSS)
are designed as \emph{dark archives} which envisage user access
only if exceptional circumstances render a separate access
replica unavailable.  Similarly,  systems providing deposit for
copyright purposes,  e.g. at
the British Library~\cite{LegalDepositLibrariesAct},
and the KB~\cite{Elsevier2003},
often reassure publishers that deposit is not a
risk to their business models by placing severe restrictions
on access to the deposited copy.  For example,  they may provide access
only to readers physically at the library.

Because users access the typical preserved data item very infrequently,
the system cannot rely on user accesses to detect,
and thus trigger the response to, errors and failures.
Provision must therefore be made for
regular audits at sufficiently frequent intervals to keep the
probability of failure at acceptable levels.
Errors and failures in system components may
be \emph{latent}~\cite{Reason1990short},  that is they may only reveal
themselves long after they occur.  Even if user detection of corruption
and failure were reliable, it would not happen in time to prevent loss.

A second important reason why audit mechanisms are important is that
digital preservation systems are typically expensive to operate,  and subject
to a high risk of economic failure.  Different systems
have different business models,  but they fall into two broad groups:
\begin{description}
\item [First-party systems] which store information belonging to the
organization operating the system.  Examples are internal corporate
systems,  state archives such as DAITSS, the US National Records and
Archives Administration (NARA), and the UK Public
Records Office, and LOCKSS peers,  which operate like a paper library storing
a ``purchased'' copy of the content.
\item [Third-party systems] which provide their customers or the taxpayers
the service of holding information belonging to a publisher.  Examples
are national library copyright deposit systems,  and archives such as
JSTOR~\cite{AboutJSTOR} and Portico.
\end{description}

Whether the funds come from internal sources,  the taxpayers,  or customers
of the digital preservation service,  the funders will require evidence
that, in return for their funds,
the service of providing access is actually being provided.
Audit mechanisms are the means by which this
assurance can be provided,  and the risk of economic failure mitigated.

\subsubsection{Audit During Ingest}
\label{sec:Strategies:audit:ingest}

Some digital preservation systems (e.g. DSpace and the KB's system)
ingest content using a
push model in which the content publisher takes action to deposit
it in the system,  whereas others (e.g. the
Internet Archive~\cite{AboutInternetArchive} and the
LOCKSS system) ingest using a pull model;  they crawl the publisher's
web sites to ingest content.

Neither process is immune from the threats outlined
above.  Some form of auditing must be used to confirm the authenticity
of the ingested content~\cite{bekaert2005}.

The LOCKSS ingest process is driven by a Submission Information Package
(SIP, in OAIS terminology)
The SIP can come in one of two forms,  an OAI-PMH~\cite{vandeSompel2004}
query asking for all new content since
the last crawl,  or a \emph{manifest page} that points to starting points
for a targeted crawl.  Neither directly specifies all the files to
be collected (for example, the OAI-PMH query may miss preexisting files
that new files include by reference).
In each case further crawling is required to ensure that the content
unit is complete.
Each peer collects its replica independently;
network and server problems may cause the set of collected URLs
to differ between the peers.
The normal LOCKSS audit process
(see Section~\ref{sec:Strategies:audit:mutual}) serves
to find the differences, and resolve them by re-crawling the
publisher's web site.  The peers thus arrive at a consensus as
to what the publisher is publishing\footnote{This currently limits
the system 
to preserving content that is published once and thereafter remains
unchanged.  Work is underway to lift this restriction
although preserving sites such as BBC
News ``updated every minute of every day''~\cite{BBCNewsSite}
with complete fidelity will remain beyond the state of the art.}.

\subsubsection{Third-party Audit}
\label{sec:Strategies:audit:external}

One common approach to audit involves retrieving a sample of the system's
content,  computing a message digest of the retrieved content,  and
comparing it with a message digest of the same content computed earlier
and preserved in some way other than in the preservation system.
If the previous message digest is
computed over the entire item submitted (the SIP)
and thus includes the metadata,
and if the system is capable of retrieving the entire SIP as part or all
of a Dissemination Information Package (DIP),  this has the attractive
property of being an end-to-end validation of the system's performance.

There are a number of problems with this approach.
The first is that the content and its previous digest are both
bit strings.  A mismatch between the current digest and the previous one
means that either the content \emph{or the digest} or both are corrupt.
Both bit strings must be preserved between audits.
Ideally the digests must be preserved 
by some means of digital
preservation other than the system being audited~\footnote{
Storing the previous message digests in the \emph{same} system can
be useful,  but it does
not protect against operator error,  external attack,  internal attack
or software failures.}.  Admittedly,  they are
much smaller than the SIPs and DIPs,  but there is at least one for
every SIP.  A large system cannot,  therefore,  rely on techniques such
as printing them on acid-free paper;  locating and transcribing the
digests would be too time-consuming and error-prone.
The digests need to be stored in
a database that can be queried during the audit.  In the limit this
sets up an infinite regress of preservation systems.
So-called ``entanglement'' protocols~\cite{Maniatis2002b}
can be used to mitigate this risk by preventing attackers re-writing
history to change previous message digests without detection,  but they
are complex and have yet to be deployed in practice.

Other problems are identified by Henson~\cite{Henson2003}.  The most
important is that,  while disagreement between the current and previous
digests gives a very strong presumption that either the content or
the digest is corrupt,  agreement between them gives a much weaker
presumption that they are unchanged.  This is not just because an
attacker might have been able to change both the content and the digest,
but also because digest algorithms
are inherently subject to \emph{collisions}, in which two different
inputs generate the same digest.  Digest algorithms are designed to
make collisions unlikely,  but some of the assumptions underlying
these designs do not hold in digital preservation applications.
For example, the analysis of the algorithm normally assumes that
the input is a random string of bits,  which for digital preservation
is unlikely.

Another is that,  like encryption algorithms,  over time message digest
algorithms become vulnerable.  Recently,  for example,  the widely
used MD5~\cite{Klima2005} and SHA1~\cite{Wang2005} algorithms appear to have
been broken.  Breaks like these are,  in effect,  latent errors because there
may be considerable delay between the actual break and knowledge of
it becoming public.  During this time auditing with message digests
may be ineffective.

If a digital preservation system audits against previous message digests
it must preemptively,
before the current algorithm is broken,
replace it.
To do so, it should audit against the current
digest to confirm that the item is still good then compute
a digest using the replacement algorithm.
This should be appended to the
stored list of digests for the item.
DAITSS is an example of a system auditing against previous digests
stored in the system itself;
it uses two different algorithms in parallel to increase the
reliability of audit and reduce the risk from broken algorithms.

Note that the result of format migration will have a different digest from
the original and, if it is itself preserved,  must have its own stored
list of digests.  This is another reason why all systems we have found
preserve the original in addition to (e.g. ANA) or instead of (e.g. LOCKSS)
the migrated version (see Section~\ref{sec:Strategies:migration}).

\subsubsection{Mutual Audit}
\label{sec:Strategies:audit:mutual}

An alternate approach to audit that is not subject to the risks of previous
message digests is used in the LOCKSS system.  Instead of trying to prove
that an individual replica is unchanged since the previous digest,
the LOCKSS audit mechanism proves at regular
intervals that the replica agrees with the consensus of the replicas
at peer libraries.  An attacker seeking to change the content has,
therefore,  to change the vast majority of the replicas within a
short period.  If there are a sufficient number of independent replicas
this can be made very hard to do,  especially in the face of the system's
internal intrusion detection measures~\cite{Maniatis2005ToCS}.

The LOCKSS audit involves peers computing and exchanging message digests;
they do
not have to reveal their content to the auditor.  This has disadvantages,
in that it is not an end-to-end audit,  and advantages,  in that it
prevents the audit mechanism being a channel by which content could leak
to unauthorized readers.
A peer whose content
doesn't match the consensus of the peers can repair it from the
original publisher,  if it is still available,  or from other peers.

This mechanism does not depend on anything but the content itself being
preserved for the long term,  and is less at risk if the message digest
algorithm is broken.  Nevertheless,  a system that used both forms of
audit would be more resistant to loss and damage than either alone;
the advantages of adding previous message digests to the LOCKSS system
are outlined in~\cite{Bungale2005}.

This mechanism also has implications for format migration.  Obviously,
once the peers have reached consensus about the information ingested,
it can ensure that this consensus is preserved.  Now,  suppose that
format migration becomes necessary.  Consensus must be re-established
on the migrated version before the mechanism can be applied to it.
This is one reason why the LOCKSS system preserves only the original.

\subsection{Economy}
\label{sec:Strategies:economy}

Techniques for reducing the cost of systems are always valuable,
but they are especially valuable for digital preservation systems.
Few if any institutions have an adequate budget for digital
preservation; they must practice some form of economic triage.
They will preserve less content than they should,  or take
greater risks with it,  to meet the budget constraints.
Reduced costs of acquiring and operating the
system flow directly into some combination of more content being
preserved or lower risk to the preserved content.

We discuss cost reduction at each of the stages of digital preservation,
ingesting the content,  preserving it,  and disseminating it to
the eventual readers. At each stage we identify a set of cost components,
not all of which are applicable to all systems.

\subsubsection{Economy in Ingest}
\label{sec:Strategies:economy:ingest}

The cost of ingesting the content has three components:  the cost of
obtaining permission to preserve the content,  the cost of actually
ingesting the content,  and the cost of creating and ingesting any
associated metadata.  

\textbf{Obtaining Permission}

Under the US Digital Millennium Copyright Act~\cite{DMCA}
and similar legislation overseas,
permission from the copyright owner is required to make and preserve
copies of copyright material.  This applies equally to open access,
subscription and pay-per-view content.  Some digital preservation systems,
including internal corporate systems and University institutional
repositories such as DSpace,  are intended to preserve content
whose copyright is owned by the host institution.
They can thus assume permission,  although obtaining explicit confirmation
of this for each item ingested might be worth the cost.

The Internet Archive takes the approach that
``to ask permission is to court denial'',
collecting and preserving copyright content
without obtaining permission.  The downside of this approach is that
if any copyright owner objects the material in
question must be immediately removed,
not a viable policy for important curated
collections.  Other
systems must obtain and preserve a record of their permission to preserve.

Negotiating and obtaining this permission can be difficult,  time-consuming
and expensive.  Copyright deposit systems have an established legal
framework in which to operate~\cite{LegalDepositLibrariesAct},
and legal incentives for
publishers to cooperate,  which can greatly reduce costs.  Other systems
must negotiate individually with each publisher.  The costs of doing
so have been identified as a major impediment to preservation of
electronic journals~\cite{Cantara2003short}.

The LOCKSS system allows access to each replica only to the host
institution's readers and is thus able to use a simple one-paragraph
addition to the publisher's existing license terms.  Experience so far
has shown the cost of negotiating permission to be manageable for larger
publishers,  where one negotiation covers many journals,  but
a significant problem for smaller single-journal publishers,
such as those being selected for preservation by the LOCKSS
Humanities Initiative~\cite{LOCKSShumanities}.  If it is
necessary for each and every journal,  even a very
cheap and easy negotiation gets expensive.
Wider adoption of the Creative Commons license~\cite{CreativeCommons},
which provides the permission needed for preservation and thus eliminates
negotiation,  could greatly reduce the cost of preservation.

\textbf{Ingesting Content}

Just as with obtaining permission,
if the ingestion of content and any necessary audit to
establish its authenticity can be automated
the per-item cost of ingestion will normally be insignificant.  To the
extent to which humans are involved in the ingest process the cost
of the process can be very significant.

\textbf{Ingesting Metadata}

Much of the discussion of digital preservation has focused on the metadata
rather than the content itself,  for example on what metadata should
be preserved along with the
content~\cite{PREMIS,PMFWG,LCCoreMetadata,NLAPmeta},  and on standards
for representing it~\cite{METS,JPEG2000}.  There has been less focus
on where it comes from,  and on the impacts the costs of creating,
validating and preserving it can have on the overall economics of
the system.

To the extent to which metadata,  especially format and bibliographic
metadata, can be supplied by the original creator of the content,
or extracted automatically from the content itself~\cite{JHOVE,Bollacker1998},
the cost impact will be low.  To the extent to which content must
be elaborated by hand with metadata,  the cost impact will
be significant.  The trade-off between preserving more content,
and providing better quality of metadata for the content that is
preserved,  can be very sharp.

It should be noted that the value of hand-generated format metadata
in assisting format migration has yet to be demonstrated in production
systems,  and that in an era when access via search is dominant the
value of even high-quality bibliographic metadata is suspect.  
Requirements for hand-generated metadata not clearly based in an
intended use can easily become counter-productive~\cite{MacKenzieSmith2005a}.

The ingestion workflow implemented by DSpace collects hand-generated,
high-quality
metadata from the submitter of the content~\cite{MacKenzieSmith2004}.
But the complexity this adds to the ingest process has caused some
resistance to its adoption~\cite{WardLlonaLally,Branschofsky}.
The LOCKSS system's ingest process is completely automated,
collecting only the metadata provided by the publisher in its
web pages.  This has been criticized as inadequate~\cite{Celeste2004};
systems such as CiteSeer~\cite{Bollacker1998} and Google Scholar
have shown that automatic extraction of metadata can be effective
but this technology has yet to be incorporated.

\subsubsection{Economy in Preservation}
\label{sec:Strategies:economy:preservation}

The cost of preserving the content and its associated metadata
has three components: the cost of
acquiring and continually replacing the necessary hardware and software,
operational costs such as power, cooling, bandwidth, staff time and
the audits needed to assure funders that they are getting their money's
worth,  and the cost of the necessary format migrations.

Systems with few replicas have to be very careful with each of them,
using very reliable enterprise-grade storage hardware and expensive off-line
backup procedures.
Systems with many replicas can be less careful with each of them,
for example using consumer-grade hardware and depending on other replicas to
repair damage rather than using off-line backups.
Our experience is is that the per-replica cost can in this way be reduced enough
to outweigh the increased number of replicas.

\textbf{Storage}

The economics
of high-volume manufacturing mean that consumer-grade disk drives
are vastly cheaper and only a little less reliable than enterprise-grade
drives.  Based on Seagate's Mean Time To Failure (MTTF) specifications,
a 200GB consumer Barracuda
drive has a 7\% probability of failing in a 5-year service
life~\cite{SeagateBarracuda} where a
146GB enterprise Cheetah has a 3\% probability of
failing~\cite{SeagateCheetah}.
But the Cheetah costs about \$8.20/GB whereas the Barracuda costs only
about \$0.57/GB\footnote{Prices from TigerDirect.com 6/13/05}.

In addition to the severe failures predicted by the MTTF specifications,
drives specify a rate of unrecoverable bit errors,  $10^{-14}$ for
the Barracuda and $10^{-15}$ for the Cheetah.  This is a very low
probability,  but the disks contain over $10^{12}$ bits.  About one in
every 62 attempts to read every bit from a Barracuda will encounter
an unrecoverable bit error;  the corresponding figure for the Cheetah
is about 1 in 860.
The disks also transfer data very fast.  
Even if the drive averages 99\% idle,  over a 5-year service life
the Barracuda will suffer about 8 and the Cheetah about 6 unrecoverable
bit errors.

Because the in-service failure probability even for expensive drives is so
high,  enterprise storage systems use replication techniques such
as RAID~\cite{Patterson1988}.
These ``internal'' replicas are costly but of little value in
digital preservation~\cite{Baker2005a}.
They provide high availability,  but spending heavily to improve
availability is hard to justify
for systems such as dark archives where the probability of a user access
during the recovery time from a disk failure is low.
They improve the reliability of the data,  but not enough to justify
their cost.
The replicas are tightly coupled to each other and
are thus subject to many correlated failure modes~\cite{Chen1994,Talagala1999}.

Another reason why digital preservation systems might not want to
use enterprise-grade hardware is the cost of power and cooling,
which can be substantial over the long lifetime of the system.
Enterprise hardware has to meet exacting performance targets and
typically does so by using power extravagantly.  Preservation systems
have much lower performance targets and can save power both by
using consumer-grade hardware and by under-clocking it.
The Internet Archive has led the way in engineering low-power
storage systems in this way,  spinning off a company called
Capricorn Technologies to build them~\cite{CapricornTechnologies}.

\textbf{Operation}

As with any activity involving humans,  system administration is expensive
and error-prone.  Yet digital preservation
requires very low rates of system administration error over very long
periods of time.
The obvious technique is to assign each replica to
its own administrative domain,  so that a single administrative error
can affect at most one replica.
In a peer-to-peer system, such
as LOCKSS, this is naturally the case; other distributed
architectures may require more costly measures to achieve
separate administrative control
of each replica.

Attempts are sometimes made to reduce the visible cost
of system administration by running the digital preservation system
as one of a large number of services offered by a large shared
server,  or as one of a large number of services sharing a
storage infrastructure such as the
Storage Resource Broker~\cite{StorageResourceBroker}.
This is often a false economy.
Layering systems in this way adds significant complexity and
introduces many failure modes,
including hardware, software, network, operational and administrative
failures, that are absent or much less significant in dedicated systems.
These add greatly to the risks to the stored content.
In particular,
it is impossible to prevent errors in other systems which share
the infrastructure but are unrelated to digital
preservation damaging preserved
content.  Machine and
administrative boundaries can be very
effective at preventing faults propagating.

The only approach to reducing operational costs while maintaining
low rates of operator error is to eliminate,  as far as possible,
the system's need for operator intervention.  The large number of
replicas envisaged for the LOCKSS system forced it to adopt this
``network appliance'' approach~\cite{Rosenthal2003b},  which has
been successful in making the per-replica cost of administration
affordable.

\textbf{Format Migration}

Format migration involves both engineering costs,  in implementing the
necessary format converters,  and operational costs,  in applying them
to the preserved content.  The engineering costs will be equivalent
whatever approach is taken,  but the operational costs will vary.
The operational cost of batch migration may be large and will be
incurred at unpredictable intervals,  making it difficult to budget.
This raises the specter of economic triage,  discarding material
whose migration cost exceeds its perceived value.  The operational
costs of the LOCKSS approach of transparent on-access migration are
minimal~\cite{Rosenthal2005a}.

\subsubsection{Economy in Dissemination}
\label{sec:Strategies:economy:dissemination}

The cost of disseminating the content has two components;  the cost
of complying with any access restrictions imposed by the agreement under
which the content is being preserved (see
Section~\ref{sec:Strategies:economy:ingest}),  and the cost of actually
supplying copies to authorized readers.

Complying with the access restrictions typically involves
an authentication system; Shibboleth~\cite{Shibboleth}
is a current example.
This is a system of comparable complexity to the digital preservation
system itself which must be adopted,  maintained,  audited and replaced
with a newer system as it becomes obsolete.  There are administrative
costs involved too,  as users are introduced to and removed from the
system,  and as the publishers with whom the agreements were made
need reassurance that they are being observed.

Actual dissemination costs such as the cost of operating a web server
and the bandwidth it uses are likely to be relatively low,  given
the archival nature of the preserved content.  Content that is expected
to be popular,  such as the UK census data~\cite{Richardson2002},
will typically be
disseminated from the preservation system once to form a temporary access
copy on an industrial-strength web server.

\subsection{Sloth}
\label{sec:Strategies:sloth}

Digital preservation is almost unique among computer applications
in that speed is neither a goal nor even an advantage.  There is
normally no hurry to ingest content,  and no large group of
readers impatient for it to be disseminated (see
Section~\ref{sec:Strategies:audit}).  As described above,
the lack of a need for speed can be leveraged to reduce the cost of
hardware,  power and cooling.  It can also reduce the cost of
system administration by increasing the window during which
administrator response is required.  Tasks that can be scheduled
flexibly and well in advance are much cheaper than those requiring
instant action.  But the most important reason for sloth is that
a system that operates fast will tend to fail fast,
especially under attack~\cite{Williamson2002,Forrest2000}.
Slow failure,
with plenty of warning during the gradual failure,  is an
important attribute of digital preservation systems,  as it
allows time for recovery policies to be implemented before
failure is total.

The LOCKSS system is an example of the sloth strategy.  Its design
principle of running \emph{no faster than necessary} was sparked
by an early talk by Stewart Brand and Danny Hillis about how the
same principle applied to the design of the ``Clock of the Long
Now''~\cite{LongNowClock}.  The principle is implemented by
rate-limiters,  which apply among other things to collecting content
by crawling the publisher's web site,  so as not to compete with
actual readers,  and the audit mechanism,  so as to prevent an
attacker changing many replicas in a short period.

\section{Requirements}
\label{sec:Requirements}

Digital preservation systems have a simple goal,
that the information
they contain remain accessible to users over a long period of time.
In addressing this goal they are subject to a wide range of threats,
not all of which are relevant to all systems.  We have
also shown a wide range of strategies,  each of which is used by
at least one current system.  But the various systems use various
techniques to implement each strategy.

The failure of a digital preservation system will become
evident in finite time,  but its success will forever remain
unproven.  Given this,  and the diversity of
threats and strategies,  it seems premature to be imposing
requirements in terms of particular technical approaches.
Rather,  systems should be required to disclose their solutions
to the various threats,  and other aspects of the strategies
they are pursuing.  This will allow certification against a
checklist of required disclosures,  and allow customers to
make informed decisions as to how their digital assets may
most economically reach an adequate level of preservation
against the threats they consider relevant.

Here is the list of suggested disclosures our bottom-up process
generated:

\begin{enumerate}
\item Systems should have an explicit threat model,  disclosing
against which of the threats of Section~\ref{sec:Threats} they are
attempting to preserve content,  and how they are addressing
each threat.
\item Systems should disclose how their replicas are created and
administered,  and how any damage is detected and repaired.
\item Systems should disclose the policies and mechanisms they
implement to protect intellectual property. Specifically:
\begin{itemize}
\item If a system is intended to hold only material whose copyright
belongs to the host institution,  it should disclose how it assures
that this is in fact the case.
\item If a system is intended to hold material whose copyright
belongs to others,  it should disclose information about the
agreement under which it is held,  such as whether and under what
terms the agreement can be revoked by the copyright holder,
and how the permission granted is verified,  recorded as metadata
and preserved.
\item If a system is intended to hold material not covered by
copyright, such as US government documents within the US,
it should disclose how it assures that this is verified,  recorded as metadata
and preserved.
\end{itemize}
\item Systems should disclose their external interfaces,  in particular
their SIP and DIP specifications.  They should disclose whether,
to assist external auditing,  they are capable of disgorging a DIP
identical to the SIP that caused the
content in question to be stored,  including not just
the content but also all the metadata originally provided
(and none of the metadata that it subsequently acquired).
\item Systems should disclose their source code access
policy,  and how their source code is to be preserved.
\item Systems should disclose who will conduct audits,
how they will be conducted,  and to whom the results will
be provided.
\item Systems should disclose their policy for handling incidents of
data loss. To whom are such incidents reported and in what form
(See Section~\ref{sec:Strategies:transparency})?
\end{enumerate}

The work underway to add certification requirements
to OAIS is proceeding along similar lines,  but from a top-down
perspective~\cite{RLGCert2005}.  We note that, while there are
strong relationships between the criteria in the current draft
of these requirements and our suggested disclosures,  there are
very few exact correspondences.

We hope this list will help the process of coming to consensus on
a set of requirements for systems to be certified under the OAIS
standard.  We also hope that it will assist system designers and
authors of papers about systems by providing a checklist of
topics which they have surely considered,  but which they may
have considered too obvious to
document~\cite{jantz2005,vanVeen2005,hyatt2005,senserini2005}.

\section{Acknowledgments}
\label{sec:acknowledgments}

We are grateful to
Bob Sproull,
Michael Lesk,
Richard Boulderstone,
Clay Shirky,
Geneva Henry,
Brian Lavoie,
Helen Hayes
and Chris Rusbridge for
valuable comments on earlier drafts.
Mary Baker, Petros Maniatis, Mema Roussopoulos,
and TJ Giuli helped greatly with multiple reviews.

This work is supported by the National Science Foundation (Grant No.\
9907296), by the Andrew W. Mellon Foundation,
and by libraries and publishers through the LOCKSS Alliance.
Any opinions, findings, and conclusions or
recommendations expressed here are those of the authors and do not
necessarily reflect the views of these funding agencies.

{\footnotesize \bibliographystyle{acm} \bibliography{bibliography}}


\end{document}